\documentclass[twoside,12pt]{article}
\usepackage{graphicx}

\def\PLB#1#2#3{{\em Phys. Lett.} B{#1} (#3) #2}
\def\NPA#1#2#3{{\em Nucl. Phys.} A{#1} (#3) #2}

\def\PRL#1#2#3{{\em Phys. Rev. Lett.} {#1} (#3) #2}
\def\PRC#1#2#3{{\em Phys. Rev.} C{#1} (#3) #2}

\def\JPG#1#2#3{{\em J. Phys.} G{#1} (#3) #2}

\newcommand{\be}{\begin{eqnarray}}
\newcommand{\ee}{\end{eqnarray}}

\newcommand{ \la }{\langle}
\newcommand{ \ra }{\rangle}

\newcommand{ \bp }{{\bf p}}
\newcommand{ \bP }{{\bf P}}

\newcommand{ \bq }{{\bf q}}
\newcommand{ \eps }{\varepsilon}
\newcommand{ \mean }[1]{\la #1 \ra}

\newcommand{ \psirp }{\Psi_{RP}}
\def\snn{\mbox{$\sqrt{s_{_{\rm NN}}}$}}   
\def \phia {\phi_a} 
\def \phib {\phi_b}

\def\pt{p_T}
\def\vtwotwo{v_2\{2\}}
\def\vtwofour{v_2\{4\}}

\begin{document}

\title{ \vspace{1cm} Collective phenomena in ultra-relativistic
  nuclear collisions: anisotropic flow and more}
\author{Sergei A. Voloshin
\\
Wayne State University, Detroit, Michigan}
\maketitle

\begin{abstract} 
Many features of multiparticle production in ultra-relativistic
nuclear collisions reflect the collision geometry and other collision
characteristics determining the initial conditions.  As the initial
conditions affect to a different degree all the particles, it leads to
truly multiparticle effects often referred to as anisotropic
collective flow.  Studying anisotropic flow in nuclear collisions
provides unique and invaluable information about the system evolution
and the physics of multiparticle production in general. Being not able
to cover all aspects of anisotropic flow in one lecture, I decided in
the first part of the lecture to discuss briefly a few important and
established results, and in the second part, to focus, in a little
more detail, on one recent development -- a recent progress in our
understanding of the role of fluctuations in the initial conditions.
I also discuss some future measurements that might reveal further
details of the multiparticle production processes.
\end{abstract}

\section{Introduction}

In ultra-relativistic nuclear collisions, the particle momentum
distributions, and in particular particle azimuthal distributions
strongly depend on the initial geometry of the collision. Such
dependence is usually discussed in terms of anisotropic collective
flow, for a recent review, see~\cite{Voloshin:2008dg}.  Anisotropic
flow has been studied since the first experiments at Bevalac, but the
first measurement of the {\em in-plane} elliptic flow
($v_2=\mean{\cos[2(\phi-\psirp)]}>0$, where $\psirp$ is the reaction
plane angle)) by the E877 Collaboration in Au+Au collisions at the BNL
AGS~\cite{Barrette:1994xr,Barrette:1996rs} 
marked a new page in the history of
anisotropic flow as a start of flow studies in truly
ultra-relativistic collisions.  Since then, the anisotropic flow
measurements became one of the most productive direction in our
pursuit to understand the physics of the high energy nuclear
collisions and multiparticle production in general.

This lecture consists of two parts. For the first part, out of several
major results from RHIC, I have selected three 
well established and very important
measurements that are directly related to the anisotropic flow:
elliptic flow of charged particles, the number of constituent
quark (NCQ) scaling of elliptic flow, and charge dependent correlation
measurements sensitive to the so-called chiral magnetic effect (CME). 
I also compare
these RHIC results to recent LHC measurements. The selection of these
three topics is based on the following. The charged particle elliptic
flow measurement was crucial in building of the picture of sQGP -
strongly interacting/coupled quark-gluon plasma, with a conclusion
that the system created at RHIC behaves as almost ideal liquid with
the lowest ever observed viscosity over entropy density ratio.  The
NCQ scaling of elliptic flow is considered as a strong evidence for
deconfinement, one of the major characteristics of QGP.  The charge
dependent azimuthal correlations relative to the reaction plane are
directly sensitive to the CME -
the charge separation along the strong magnetic field of colliding
nuclei that is intimately related to the properties of the QCD vacuum.
The level of uncertainty in the interpretation of the above mentioned
three measurements (and the corresponding backgrounds) is different -
increasing from the first one to the last, but taking into account the
importance of the questions, the coalescence hadronization picture in
the case of NCQ scaling, and, possibly, the first direct measurement
related to the non-perturbative structure of QCD vacuum in the last,
all three measurements can not be undervalued.

The second part of this lecture is about the rapid development over
the last couple years in our understanding of the role of anisotropic
flow fluctuations, and their relation to seemingly unrelated phenomena
observed via two-particle correlations, such as the ``ridge'' and
``Mach cone''. The uncovering of this relationship has resolved
a long standing issue of the relative importance of flow fluctuations
and the so-called nonflow.  For several years we were puzzled by a
possibility to explain the difference between two- and multi- (more
than two) particle measurements of elliptic flow either almost
entirely by flow fluctuations, or again, almost entirely by nonflow
assessed via analysis of two-particle correlations. It appears that
the resolution of this puzzle is simple: two approaches are just
different descriptions of the same phenomena - the system response to
the fluctuating initial conditions. This understanding allows us to
propose totally new and promising new insights 
measurements, such as the femtoscopic analysis of
the particle production relative to the higher harmonic event planes.

\section{Major results from RHIC era and comparison to LHC data}
\subsection{Integrated and differential elliptic flow of charged particles}

\begin{figure}[h!]
\includegraphics[width=75mm]{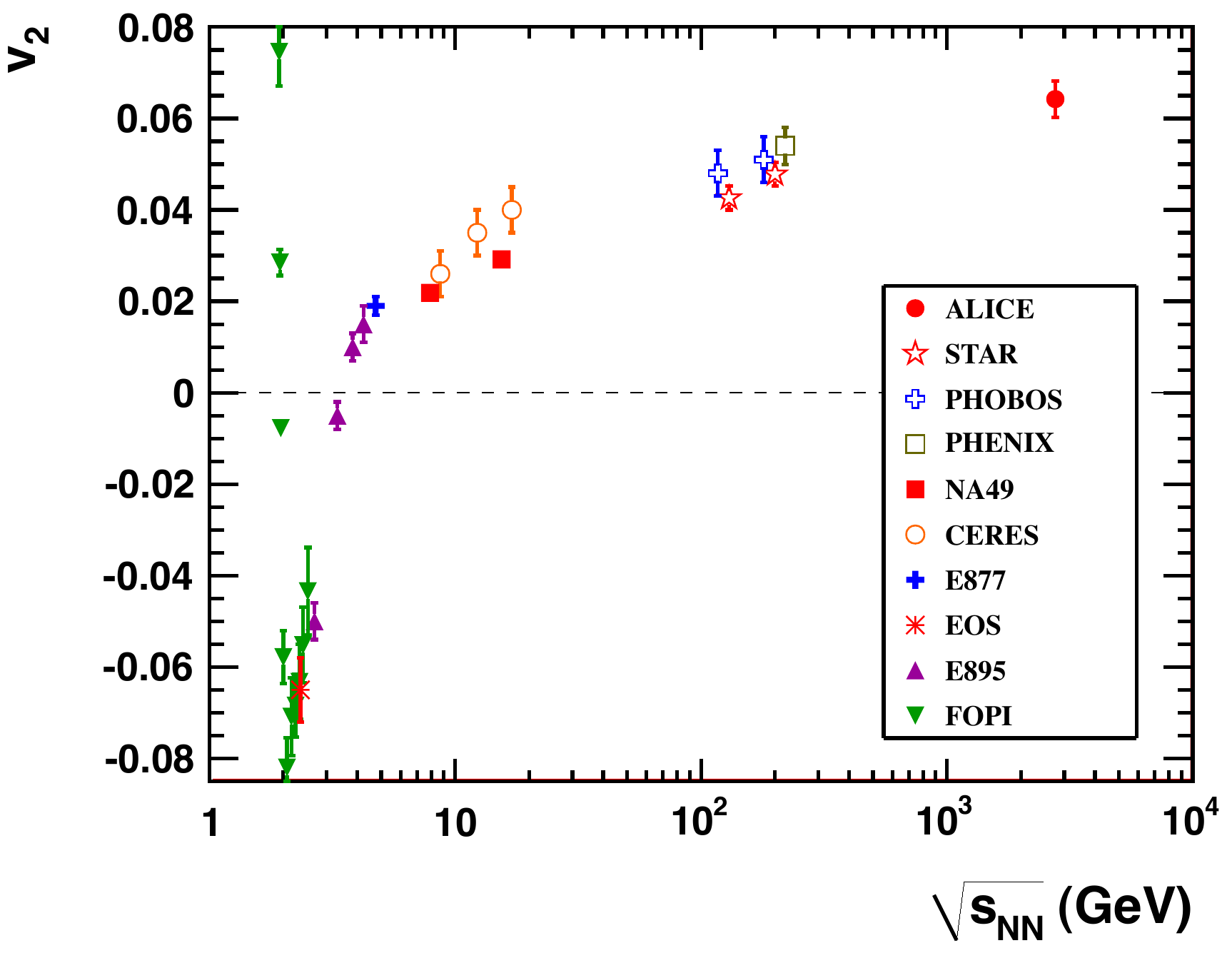}
\includegraphics[width=88mm]{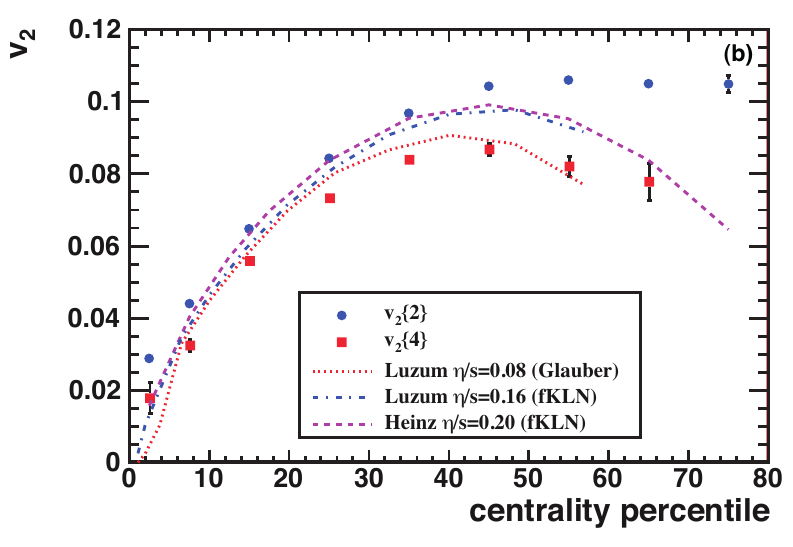}
\caption{ (Left) Energy dependence of the average elliptic flow in
  20-30\% centrality bin at different collision energies
  (from~\cite{Aamodt:2010pa}).  (Right) $v_2$ centrality dependence
  in Pb+Pb collisions at
  2.76~TeV~\cite{Aamodt:2010pa} compared to viscous hydro calculations
  (this figure is taken from~\cite{Snellings:2011sz}).  }
\label{fig:v2edep} 
\end{figure}

One of the main (and probably most widely known) result from RHIC is
the observation of strong elliptic flow ($v_2$), which for the first
time is quantitatively close to the predictions of ideal
hydrodynamics.  Although for many this result appeared as totally
unexpected, an analysis of the BNL AGS and CERN SPS data indicated
that the elliptic flow exhibited a strong increase with collisions
energy in that energy domain.  A simple extrapolation to RHIC (and the
LHC) energies, e.g. performed using suggested
in~\cite{Voloshin:1999gs} $v_2/\eps$ scaling with particle density
(here $\eps$ is the eccentricity of a nuclear collision overlap zone)
suggested a significant increase in $v_2$ up to the RHIC energies with
some kind of a saturation at lager energies. The experimental
measurements~\cite{Ackermann:2000tr} agree well with such a
conclusion, see the excitation function of elliptic flow in
mid-central collisions in Fig.~\ref{fig:v2edep}.  The recent data from
LHC~\cite{Aamodt:2010pa} supports the picture indicating that at even
higher system temperatures at LHC, the elliptic flow remains to be
large in an agreement with hydrodynamic calculations.


Ever more precise measurements of anisotropic flow was developing
along with a significant progress in viscous relativistic
hydrodynamics calculations. The comparison of the experimental results
with theory allowed unprecedented extraction of transport properties
of sQGP, first of all the ratio of shear viscosity to entropy
density. That appeared to be the lowest ever observed, of the order of
a few times the lower bound $1/4\pi$, see Fig.~\ref{fig:v2edep}, right
panel, which shows a comparison of $v_2$ centrality dependence to
different viscous hydrodynamic
calculations~\cite{arXiv:1011.5173,arXiv:1108.5323}.

\subsection{NCQ scaling of elliptic flow}

The QGP is a thermalized and deconfined QCD matter. To test if the
deconfinement has been reached in nuclear collisions is not a simple
task, in particular because at present we do not have any theoretical
model that would self-consistently describe the hadronization
process. This is the reason why the idea~\cite{Voloshin:2002wa}, that
in heavy ion collisions there might be a region in transverse momentum
where the particle production will be dominated by quark coalescence
and, at the same time, can be described by the standard coalescence
formalism attracted a lot of attention. (Note that in general the
standard coalescence formalism is applicable only to the so-called
rare processes in which the ``parent'' distributions are weakly
affected by the coalescence process -- just recall that the formalism
was developed for the light nuclei production.) Based on the
coalescence hadronization picture one can make predictions on the
dependence of the particle mean transverse momentum on particle
density, baryon-to-meson ratio, and most importantly for this
discussion, for anisotropic flow of baryons and mesons.

The essence of the coalescence formalism is the statement that the
invariant spectrum of produced particles is proportional to the
product of the invariant spectra of constituents:
\be \frac{dN_B}{d^2 p_\perp}(\bp_\perp) = C_B \left[
  \frac{dN_q}{d^2 p_\perp}(\bp_\perp/3) \right]^3, \;\;\;
\frac{dN_M}{d^2 p_\perp}(\bp_\perp) = C_M \left[ \frac{dN_q}{d^2
    p_\perp}(\bp_\perp/2) \right]^2 \ ,
\label{eq:coal}
\ee 
where the coefficients $C_M$ and $C_B$ account for the probabilities
for $q\bar{q}\to meson$ and $qqq \to baryon$ coalescence.  The quark
spectra are not uniform relative to the reaction plane, and according
to Eq.~\ref{eq:coal} the
anisotropy should be amplified almost a factor of two in meson spectra and
about factor of three in baryon spectra.  Within that picture (and
only in the limited region where the formalism is applicable),
elliptic flow, $v_2(\pt)$, of baryons is expected to be about 3/2
times larger than that of mesons~\cite{Voloshin:2002wa,Molnar:2003ff}.
Such a relationship has been observed experimentally later within
about 10-20\% accuracy -- typical for such kind of predictions (of
``additive quark model''). This observation is considered as a strong
evidence that the system created in heavy ion collision is in a {\em
  deconfined} state during most of its evolution. The preliminary
results from LHC seems to be in agreement with the NCQ scaling,
though the detailed analyses of LHC data are ongoing.

\begin{figure}[h!]
\includegraphics[width=90mm]{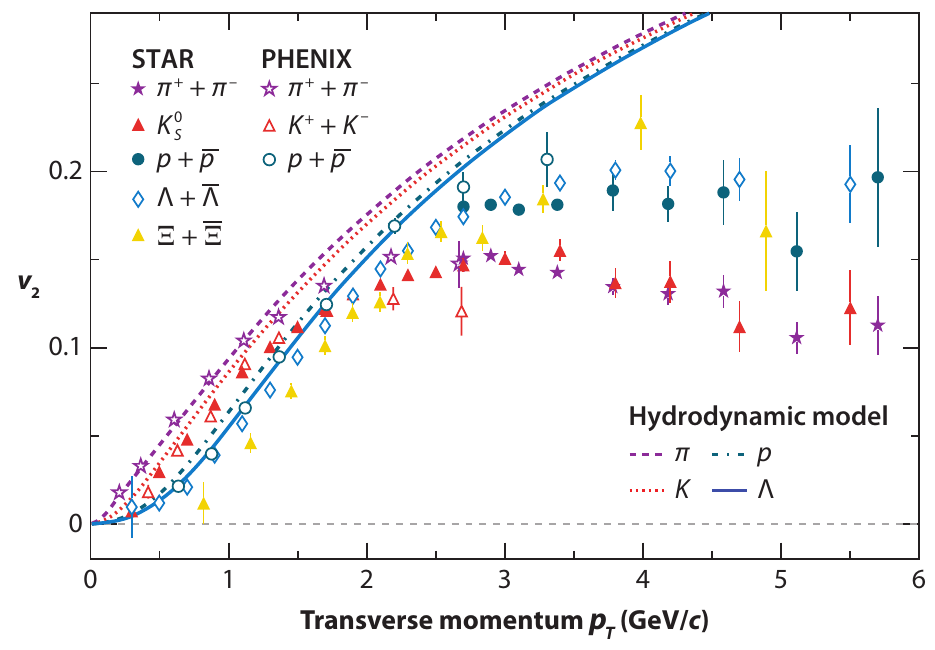}
\includegraphics[width=80mm]{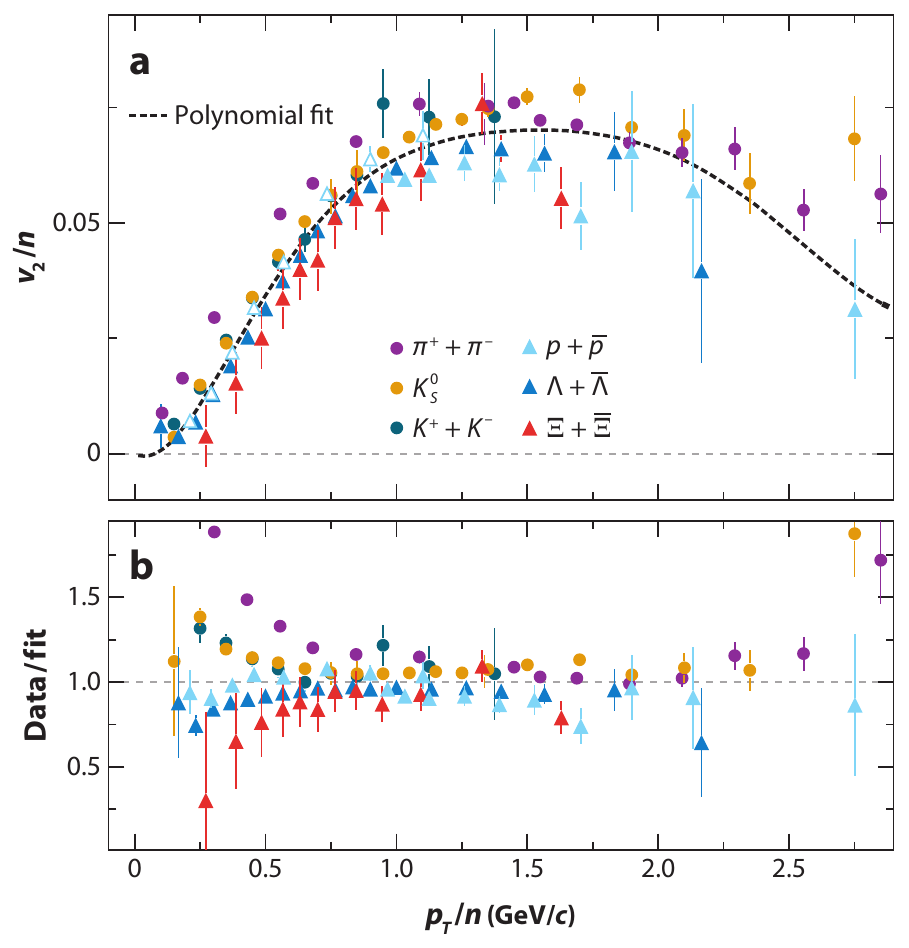}
\caption{
(Left) Identified particle elliptic flow in Au+Au collsions at
  \snn=200 GeV. (Right) Test of the NCQ scaling. 
Both figures are taken from review~\cite{arXiv:0807.4939}}
\label{fig:ncq} 
\end{figure}

\subsection{Testing the chiral magnetic effect}

\begin{figure}[h!]
\includegraphics[width=80mm]{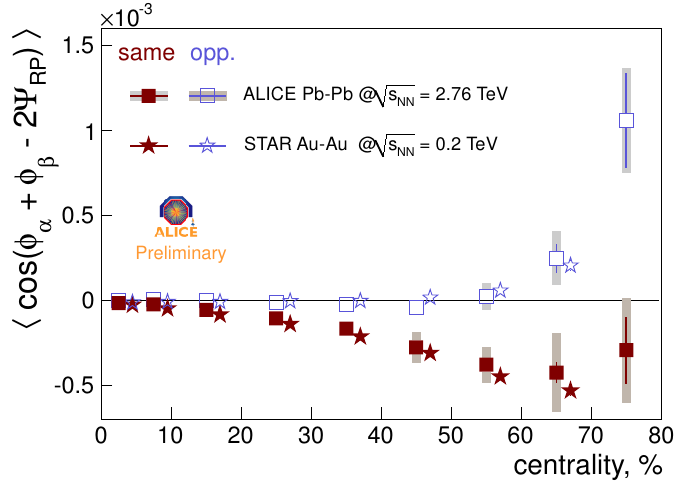}
\hspace{1.5cm}
\includegraphics[width=75mm]{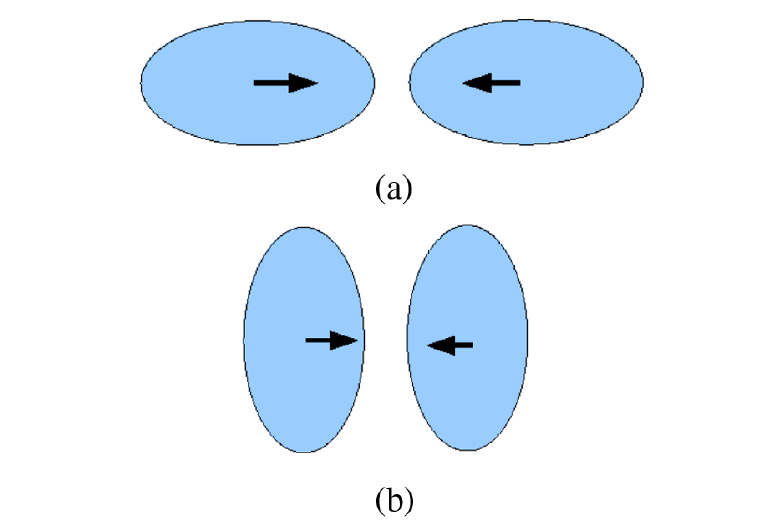}
\caption{
(Left) The correlator Eq.~\ref{eq:obs1} measured by STAR and ALICE
collaborations~\cite{Selyuzhenkov:2011xq}. (Right) Tip-tip (a) and
body-body (b) configurations of $U+U$ central collision.
}
\label{fig:cme} 
\end{figure}

QCD links chiral symmetry breaking and the origin of hadron masses to
the existence of topologically nontrivial classical gluonic fields,
instantons and sphalerons, describing the transitions between the
vacuum states with different Chern-Simons numbers.  Quark interactions
with such fields change the quark chirality and are $\cal P$~and $\cal
CP$~odd.  Though theorists have little doubt in the existence of such
fields, they have never been observed directly, e.g. at the level of
quarks in the deep inelastic scattering.  The experimental search for
the local strong parity violation in heavy ion collisions was greatly
intensified once it was noticed~\cite{Kharzeev:2004ey,Kharzeev:2007tn}
that in noncentral nuclear collisions it could lead to the asymmetry
in the emission of positively and negatively charged particle
perpendicular to the reaction plane.  Such a charge separation is a
consequence of the difference in the number of quarks with positive
and negative helicities positioned in the strong magnetic field ($\sim
10^{15}$~T) of a noncentral nuclear collision, the chiral
magnetic effect~\cite{Kharzeev:2004ey,Kharzeev:2007jp}.  The
direction of the charge separation varies in accord with the sign of
the domain topological charge, which makes the observation of the
effect experimentally difficult.  A correlator sensitive to the CME
was proposed in Ref.~\cite{Voloshin:2004vk}:
\be
\hspace*{-2cm}
 \mean{ \cos(\phia +\phib -2\psirp) } &= 
&\mean{\cos\Delta \phia\, \cos\Delta \phib} 
-\mean{\sin\Delta \phia\,\sin\Delta \phib}
\label{eq:obs1}
\\ 
& =&
[\mean{v_{1,\alpha}v_{1,\beta}} + B_{in}] - [\mean{a_{1,\alpha} a_{1,\beta}}
+ B_{out}]
 \approx - \mean{a_{1,\alpha} a_{1,\beta}} + [ B_{in} - B_{out}],
\nonumber 
\ee
where the $a_1$ coefficients describe the (first harmonic) up-down
asymmetry in particle production. Subscript $\alpha$ denotes the
particle type.  STAR Collaboration
measurements~\cite{:2009uh,:2009txa} of this correlator are consistent
with the expectation for the CME and can be considered as evidence of
the local strong parity violation.  Preliminary ALICE
results~\cite{Selyuzhenkov:2011xq}, see Fig.~\ref{fig:cme}, show very
similar signal.  The ambiguity in the interpretation of experimental
results comes from a possible background of (the reaction plane
dependent) correlations not related to CME.  As the detailed
quantitative predictions for CME do not yet exist, it is difficult to
disentangle different contributions.  At the same time there exist no
model which would fully describe the data in terms of ``conventional''
physics. The most notable in this respect is the
paper~\cite{arXiv:1009.4283} where the authors show that the
difference between the same sign and opposite sign correlations as
measured by STAR can be explained within a blast wave model that
includes charge conservation along with radial and elliptic flow with
parameters tuned to the data. 

Note that a key ingredient to CME is the strong magnetic field, while
all the background effects originate in the elliptic flow. This can be
used for a possible experimental resolution of the question. One
possibility is to study the effect in central collisions of
non-spherical $U$ nuclei~\cite{Voloshin:2010ut}, where the relative
contributions of the background (proportional to the elliptic flow)
and the CME (proportional to the magnetic field), should be very
different in the tip-tip and body-body type collisions (see right
panel of Fig.~\ref{fig:cme}, and differ from those in central Au+Au
collisions. The body-body $U+U$ collisions would correspond to a
strong background (elliptic flow due to the ellipsoidal shape of the
nuclear overlap zone) and small CME signal, as the magnetic field is
expected to be relatively weak in such collisions.

As discussed below, fluctuations in the initial conditions lead to
significant higher harmonic flow. This opens new possibilities for
estimates of the background in the measurements of charge dependent
correlations sensitive to the CME. The idea here is that the CME, the
charge separation along the magnetic field, should be zero (highly
suppressed) if measured with respect to higher harmonic event planes,
while the background effects due to flow should be still present,
although smaller in magnitude (according to higher harmonic flow).  An
example of such a correlator, where the CME contribution expected to
be strongly suppressed, would be:
\begin{equation}
 \mean{\cos(2\phi_\alpha+2\phi_\beta-4\Psi_4}, 
\end{equation}
where $\Psi_4$ is the fourth harmonic event plane. The value of
the background due to flow could be estimated by rescaling the correlator
Eq.~\ref{eq:obs1}.  
These  measurements will require good statistics;
detail interpretation would also need calculations of the 
magnetic field fluctuations.  
Finally, I note that many other experimentally possible
tests, in particular with identified particles, can be found
in~\cite{Voloshin:2009hr}.

\section{Recent developments: higher harmonic flow}

For a long time, the main systematic uncertainty in the interpretation
of the elliptic flow measurements was the unknown relative
contribution from flow fluctuations and nonflow -- the correlations
not related to the initial geometry. These two effects determine the
difference in $v_2\{2\}$ and $v_2\{4\}$ - elliptic flow measurements
with two and 4-particle cumulants. $\vtwotwo$ is biased toward higher
values by both, flow fluctuations and nonflow, and $\vtwofour$,
basically free from nonflow, is biased toward smaller values by
fluctuations.  It is thought that flow fluctuations are mostly due to
fluctuations in the position of the participating nucleons, the
distribution of which determines the participant plane (that is
different from the reaction plane). The first realistic estimates of
flow fluctuations~\cite{nucl-ex/0612021,arXiv:0707.4424} showed that
almost the entire difference between $v_2\{2\}$ and $v_2\{4\}$ can be
accounted for by flow fluctuations without much ``room'' left for
nonflow contribution.  A little earlier to these flow fluctuations
estimates, it was argued~\cite{Voloshin:2003ud} that in a nuclear
collision the spatial correlations of particles produced in the same
{\em nucleon-nucleon} collision in conjunction with the radial flow
lead to a narrow in azimuth and long ranged in rapidity
correlations. Such correlations were observed later experimentally and
called {\em ridge}.  In~\cite{Takahashi:2009na,Andrade:2010sd} it was
shown explicitly in the event-by-event hydrodynamical calculations
that the fluctuations in the initial density distribution that extends
over large rapidity range lead to the ridge structure in two particle
correlations (for a bit more detail discussion of this question
see~\cite{Voloshin:2011mg}).  An important and striking observation
made in~\cite{Voloshin:2003ud} was that such a mechanism with
realistic values of radial flow would lead to two-particle azimuthal
correlations, which again almost entirely explain the difference
between $v_2\{2\}$ and $v_2\{4\}$, but now not by flow fluctuations
but ``nonflow''. There was a puzzle.

\begin{figure}[tb]
\begin{center}
\begin{minipage}[b]{12.2 cm}
\includegraphics[width=4cm]{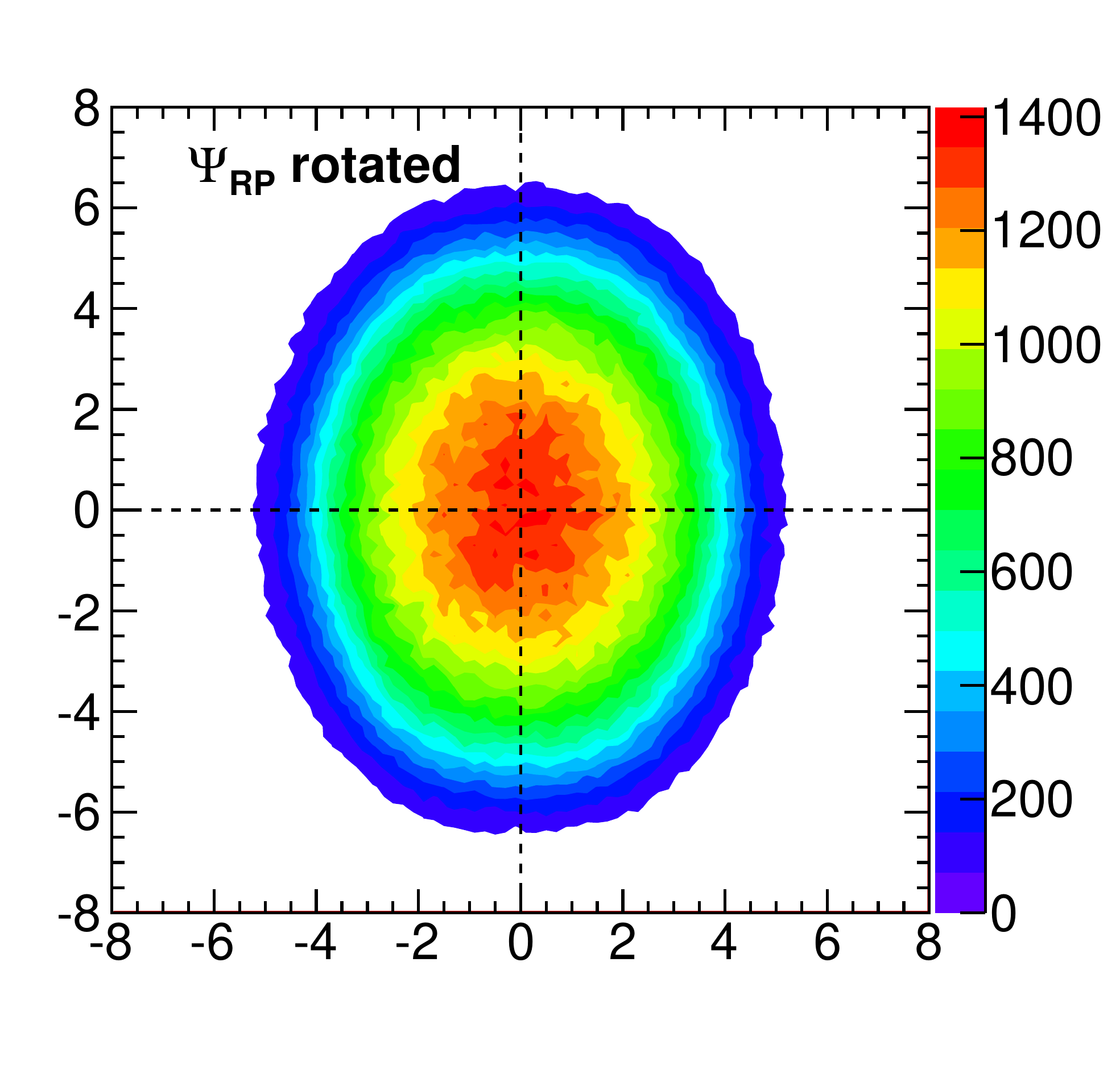}
\includegraphics[width=4cm]{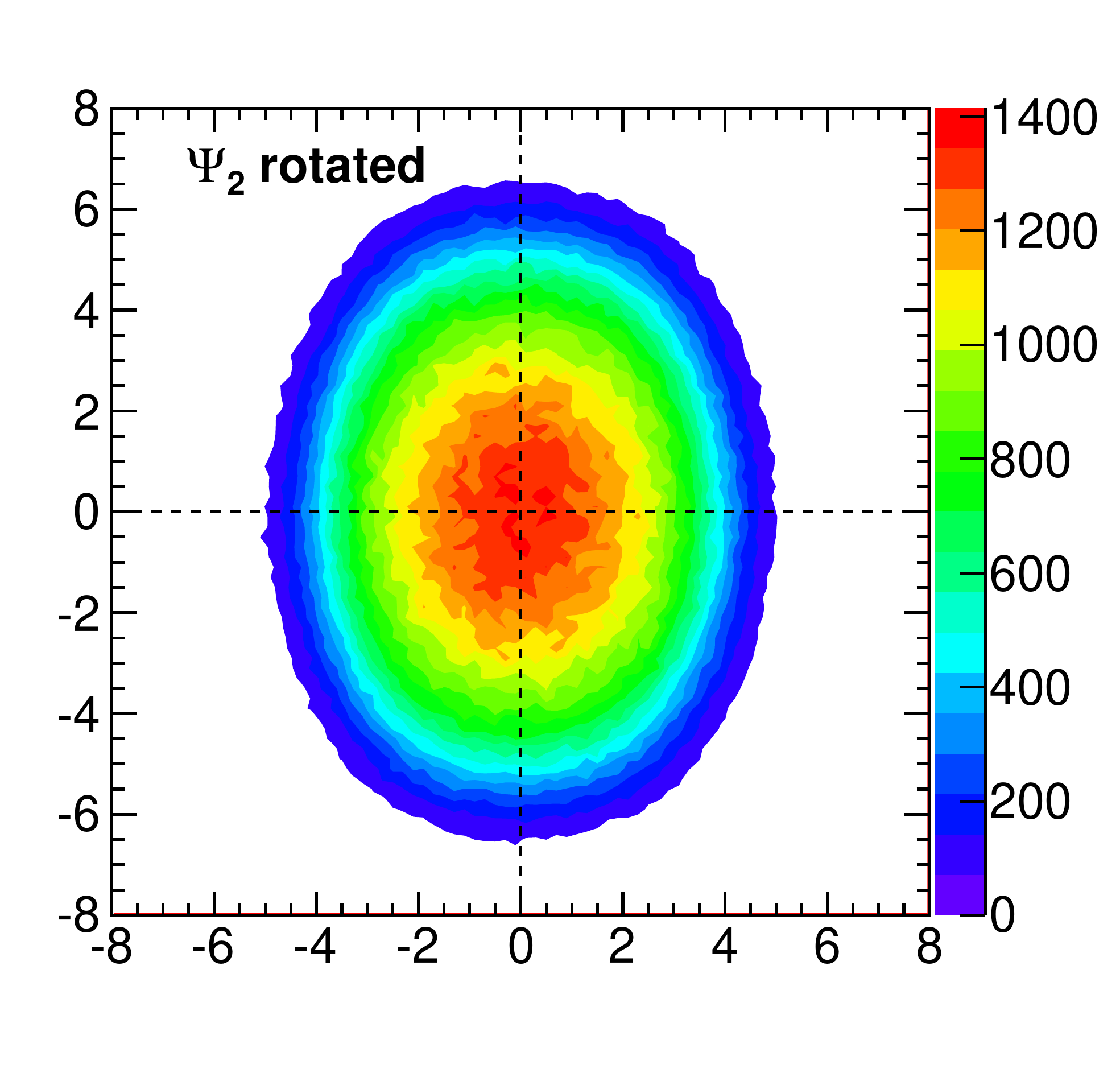}
\includegraphics[width=4cm]{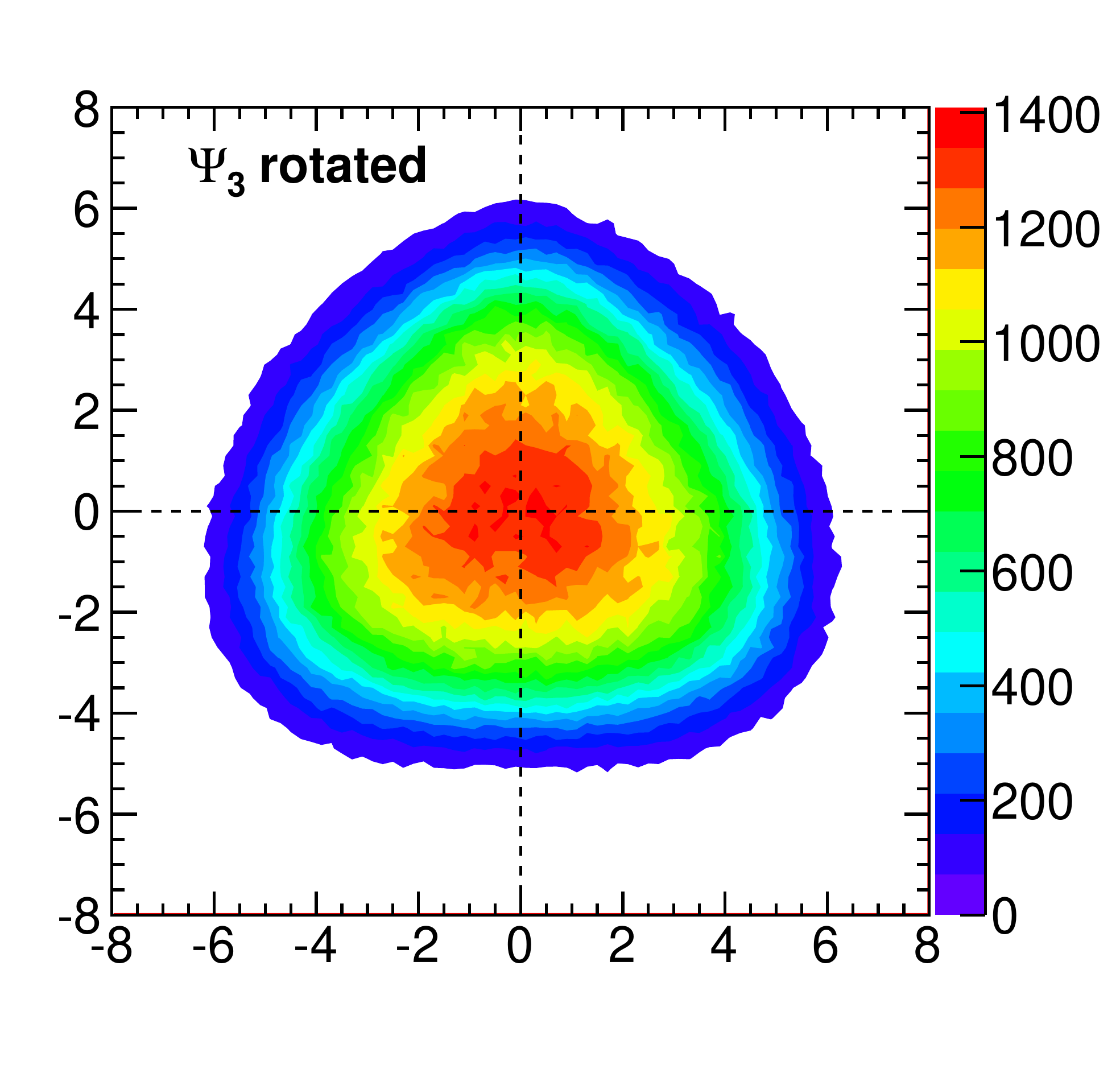}
\includegraphics[width=4cm]{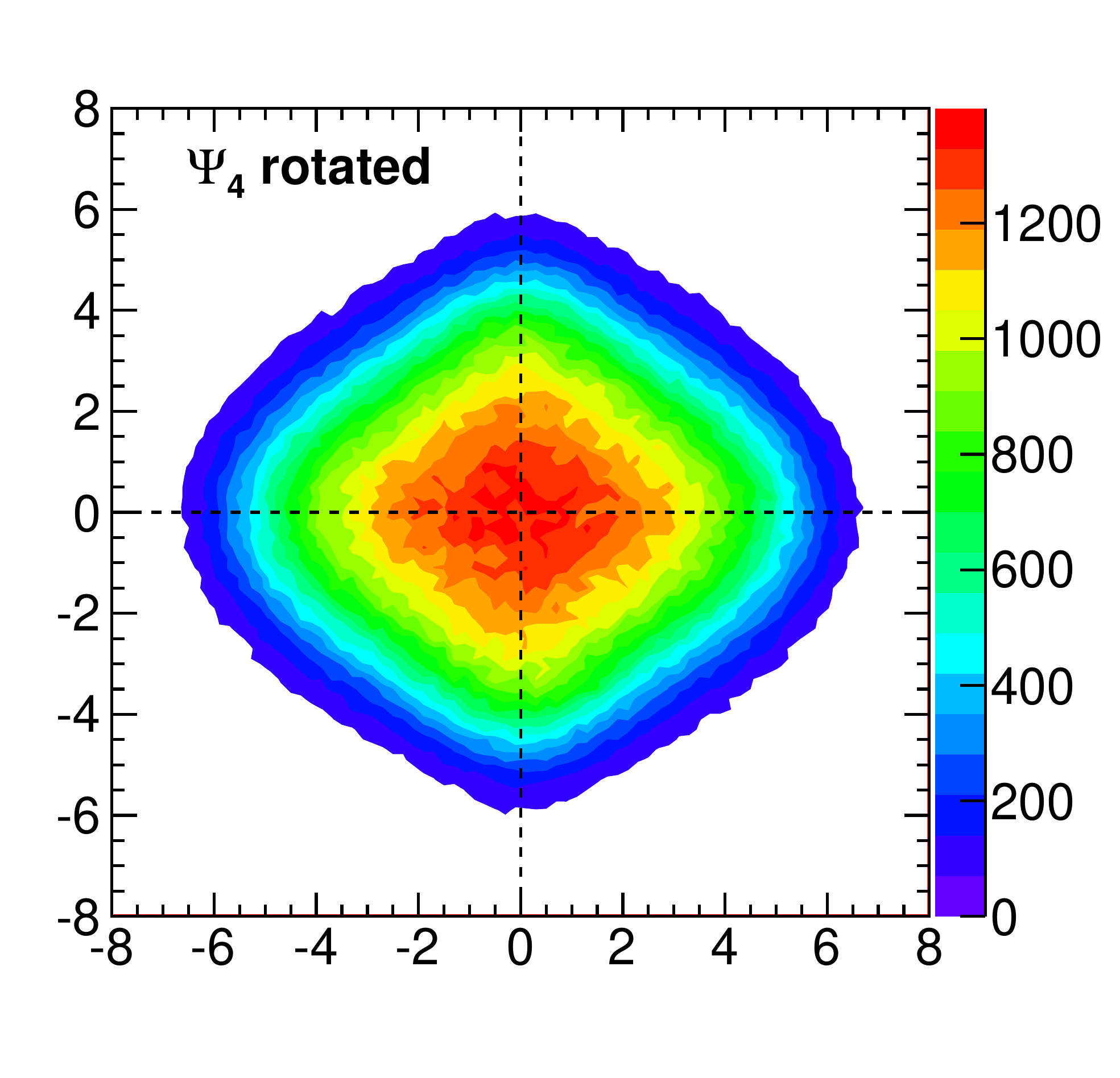}
\includegraphics[width=4cm]{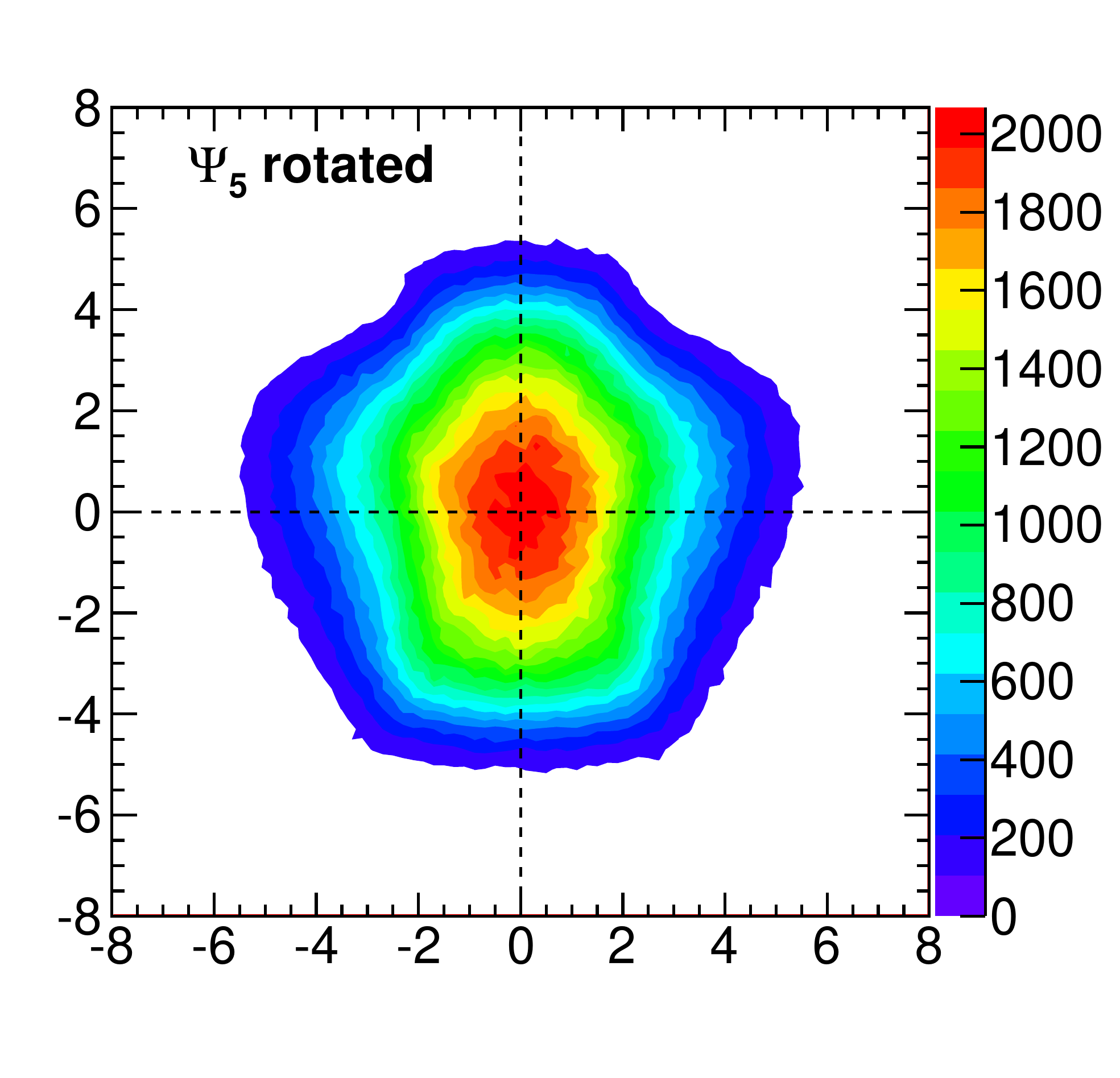}
\includegraphics[width=4cm]{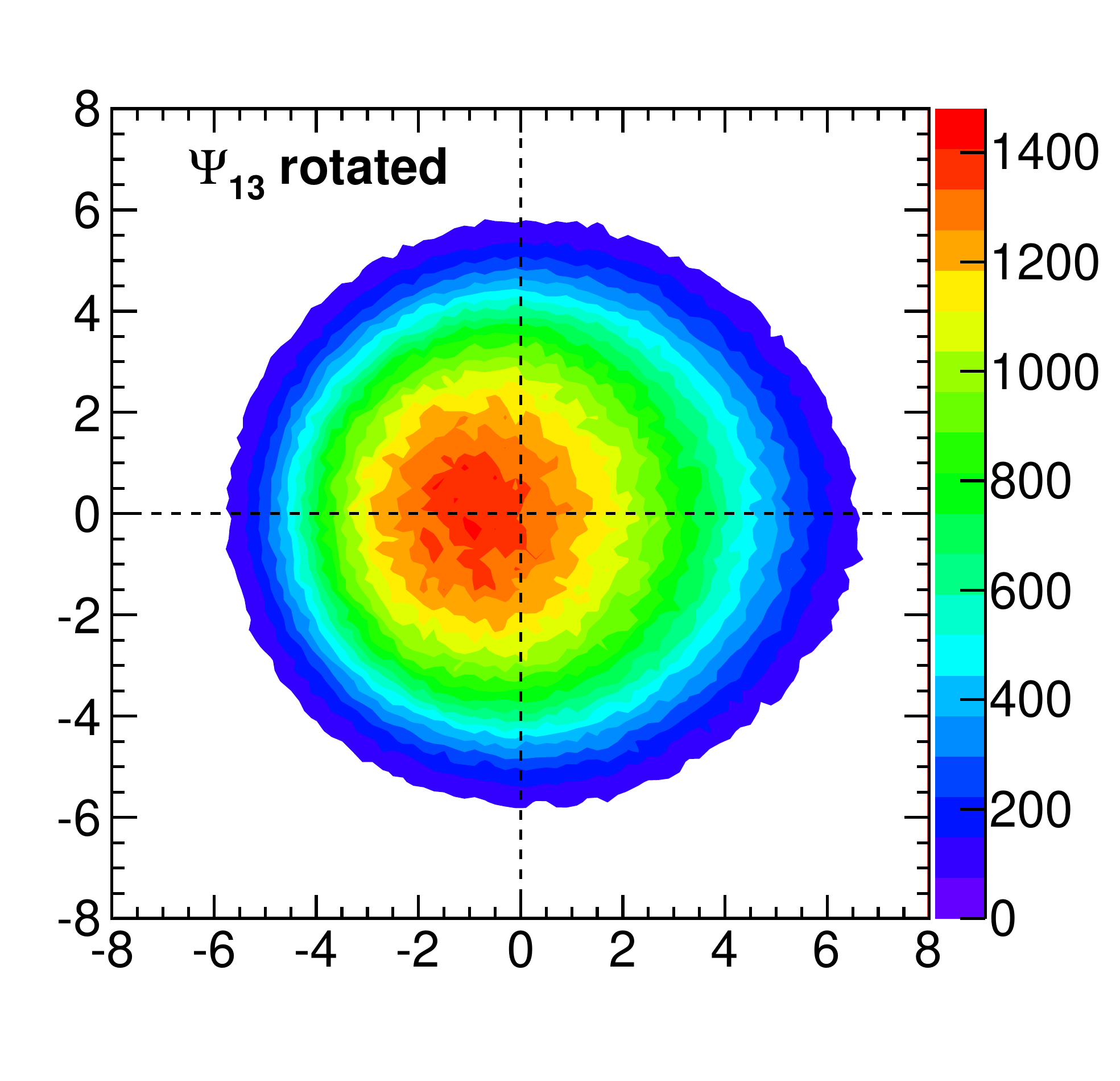}
\end{minipage}
\begin{minipage}[b]{6 cm}
\includegraphics[width=6cm,height=6cm]{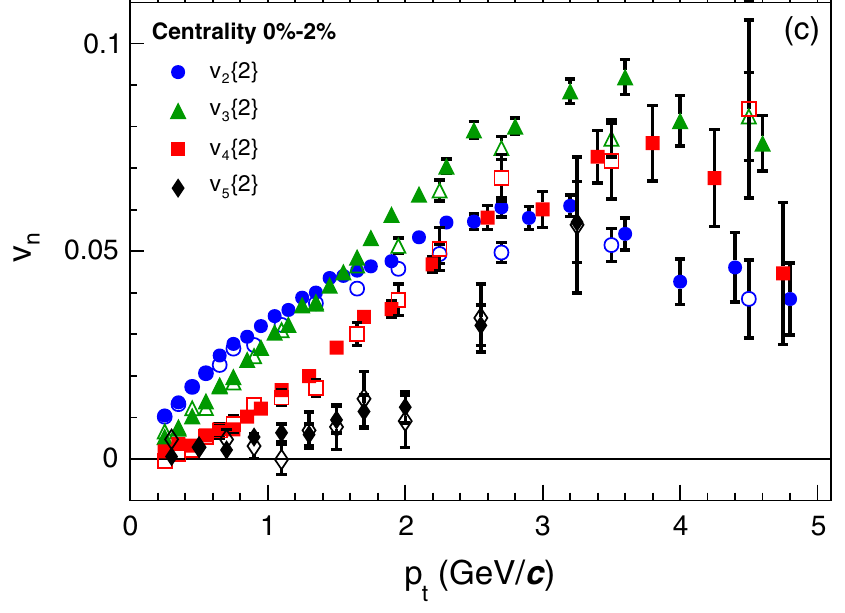}
\end{minipage}
\begin{minipage}[t]{16.5 cm}
\caption{(Left panels) 10k Au+Au collision events at $b=8$~fm rotated
  to different event planes. (Right panel) Differential flow for
  harmonics $n=1-5$ by ALICE Collaboration~\cite{Aamond:2011vk}.  }
\end{minipage}
\end{center}
\label{fig:rotated}
\end{figure}

The resolution of the puzzle appeared to be simple: although the
appearance of the ridge and anisotropic flow fluctuations look as
totally unrelated phenomena, they have the same roots and appeared to
be different descriptions of the same phenomenon -- the reaction of
the system to fluctuations in the initial density.  It was noticed
in~\cite{Mishra:2008dm,Sorensen:2010zq} that the fluctuating initial
conditions generate anisotropic flow of different harmonics.  That
followed by
understanding~\cite{Teaney:2010vd,Staig:2011wj,Shuryak:2009cy} that
the perturbations due to each of the ``hot spots'' can be treated
independently, which allowed to reformulate the problem from a
different perspective -- decompose the initial density into multipoles
and study the system response to each of the multipoles - dipole,
triangular, quadrangular, etc.  fluctuations.  Effectively, this
approach is equivalent to the rotations of each of the events to a
given harmonic symmetry plane, which results into smooth initial
conditions but with a shape corresponding to a given harmonic. To
illustrate this, in Fig.~\ref{fig:rotated} I show the participant
nucleon density averaged over the {\em same} 10k events but with each
event rotated to a correspondent harmonic event plane, identified in
the figure. These very different shapes are the origin of higher
harmonic flow that was recently measured by several collaborations at
the LHC and RHIC, see the results from ALICE
Collaboration~\cite{Aamond:2011vk} in the right panel of
Fig.~\ref{fig:rotated}.
  
Different shapes in the initial geometry of the collision, as shown in
Fig.~\ref{fig:rotated}, to a different degree will be preserved in the
system freeze-out shapes. It was shown in~\cite{Voloshin:2011mg} that
those shape can be addressed experimentally with azimuthally sensitive
femtoscopic analysis.  Recall that 
for a Gaussian source, the
correlation function appears to be a Gaussian
\begin{equation}
C(\bq, \bP) \propto 1+ \exp \sum_{i,j} R_{ij}^2 q_i q_j,
\end{equation}
where $\bq=\bp_1-\bp_2$ is the pair relative momentum,
 $\bP=(\bp_1+\bp_2)/2\approx \bp_i$ is the particle average momentum,
 and $R_{ij}^2=\mean{(r_i-V_it)(r_j-V_jt)}$ are the HBT radii.
The idea of the azimuthally sensitive femtoscopic analysis,
to study the radii dependence on the pair emission angle
 with respect to the reaction plane, was first proposed 
in~\cite{Voloshin:1995mc} with an extension 
to non-identical particle correlation in~\cite{Voloshin:1997jh}.
Details of femtoscopic analyses and discusion of the experimental results
can be found in a review~\cite{Lisa:2005dd}.

\begin{figure}[h!]
\includegraphics[width=55mm]{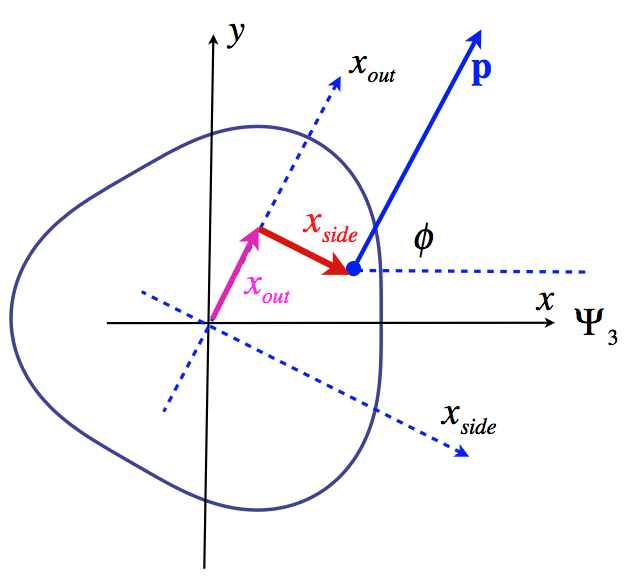}
\includegraphics[width=51mm]{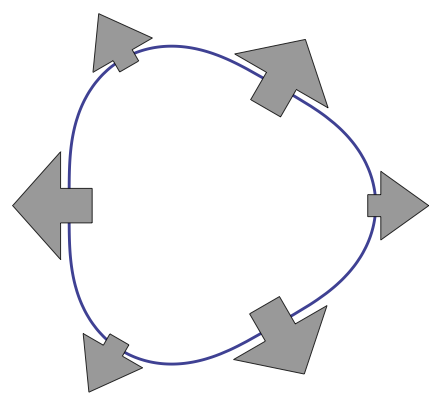}
\includegraphics[width=65mm]{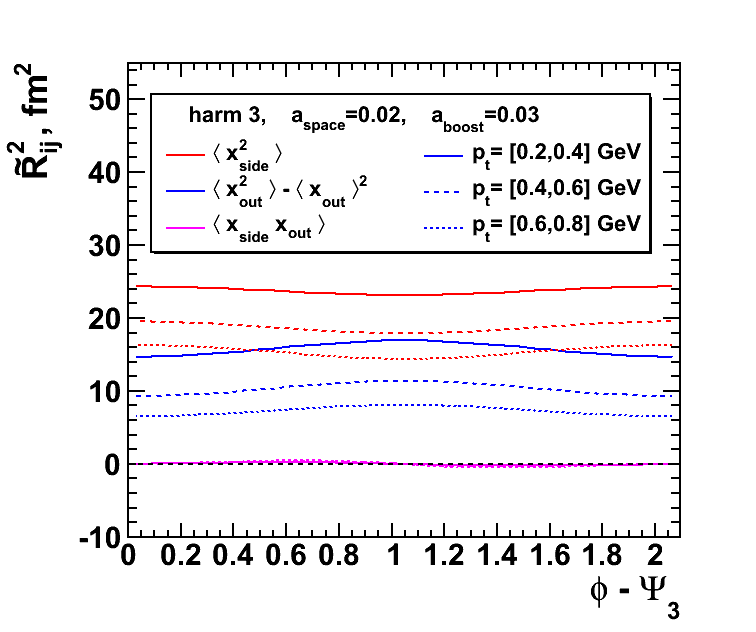}
\caption{Left: side-out coordinates. Middle: illustration
  of predominant expansion along shorter directions.
$v_n(pt)$ for typical values of parameters used in this work. 
}
\label{fig:hbt} 
\end{figure}

The dependence of the HBT radii on the higher harmonic ($n>2$) flow
appears to be a bit more complicated compared to the ``standard''
analysis with respect to the reaction plane.  In the discussion below I use a
standard side-out-long system~\cite{Lisa:2005dd} (see
figure~\ref{fig:hbt}, left).  For a {\em stationary} (not expanding)
source the radii azimuthal dependence can be expressed as
\be
R_{side}^2=\mean{x_{side}^2} = \mean{x^2}\sin^2\phi
+\mean{y^2}\cos^2\phi-\mean{xy}\sin2\phi, 
\ee 
which has only $n=2$ harmonic.  Higher harmonics azimuthal dependence
appears only as a deviation from the Gaussian shape of the correlation
function, e.g. in $\mean{x_{side}^6}$ and $\mean{x_{side}^4}$ for
triangular and quadrangular shapes respectively, which would greatly
complicate the analysis (for the effect of non-Gaussiness on HBT
radii, see~\cite{Hardtke:1999vf}).  But this is true only for a
stationary source. The picture changes if one considers azimuthal
variation in the expansion velocity, see Fig.~\ref{fig:hbt} middle
panel, where the thickness of the arrows indicate the expansion
velocity.  As shown in~\cite{Voloshin:2011mg} using a blast wave model
calculations with realistic parameters, the azimuthal dependence of
the HBT radii is significant, see Fig.~\ref{fig:hbt} right panel,
which indicates that the higher harmonics shape effects become clearly
visible and measurable.  That was also confirmed in the
AMPT~\cite{Zhang:1999bd} model calculations~\cite{Voloshin:1995mc}.

\section{Summary} 
Heavy ion collisions is unique laboratory to study QCD, including the
physics of hadronization and properties of QCD vacuum. Anisotropic
flow is one of the most important and sensitive tool in this study.
The recent progress in the understanding of the physics of anisotropic
flow fluctuations and their relation to the structures in two
particle $\Delta\eta \times \Delta \phi$ correlations, further
advances the physical interpretation of the measurement.   
Measurements with higher harmonics flow promise new insights to the
correlation measurements related to CME, and the system shape and
velocity fields via femtoscopy.

\section*{Acknowledgments}
I thank Prof. A.~Faessler and
Prof. J.~Wambach, 
the organizers of the International School of Nuclear
Physics, Erice, 2011, for the invitation to this stimulating meeting.


\end{document}